\documentclass[floatfix,aps,showpacs,twocolumn]{revtex4}
\usepackage{graphicx}
\usepackage{amssymb}

\begin{document}

\newcommand{\x}{{\bf x}}

\title{Quantum error correction of coherent errors by randomization}
\author{O.~Kern$^{(1)}$, G.~Alber$^{(1)}$, and D.~L.~Shepelyansky$^{(2)}$}
\affiliation{$^{(1)}$Institut f\"ur Angewandte Physik, Technische 
Universit\"at Darmstadt, 64289 Darmstadt, Germany\\
$^{(2)}$Laboratoire de Physique Th\'eorique, UMR 5152 du CNRS, 
Univ. P. Sabatier, 31062 Toulouse Cedex 4, France}
\date{July 30, 2004}
\begin{abstract}

A general error correction method is presented which 
is capable of correcting coherent errors originating
from static residual inter-qubit couplings in a quantum computer.
It  is based on a randomization of static imperfections 
in a many-qubit system by the repeated application of Pauli operators which
change the computational basis.
This Pauli-Random-Error-Correction (PAREC)-method
eliminates coherent errors produced by static imperfections 
and increases significantly 
the maximum time over which realistic quantum computations 
can be performed reliably. Furthermore, it does not require redundancy so 
that all physical qubits involved can be used for logical  purposes.

\end{abstract}
\pacs{03.67.Lx,03.67.Pp,05.45.Mt}

\maketitle

\flushbottom
Current developments in quantum physics demonstrate in an impressive 
way its technological potential \cite{Nielsen00}.
In quantum computation, e.g., characteristic quantum phenomena, 
such as interference and entanglement,
are exploited for solving computational tasks more efficiently 
than by classical means \cite{Deutsch,Shor,Grover97,Lloyd,Shep1}.
However, these quantum phenomena are affected easily by unknown residual
inter-qubit couplings 
or by interactions with an uncontrolled environment \cite{dephasing}.
In order to protect quantum algorithms against such undesired influences
powerful methods of error correction
have been developed over the last years. 

So far techniques of quantum error correction have concentrated 
predominantly on decoherence caused by uncontrolled
couplings to environments \cite{EC,Alber01a}. In these cases appropriate 
syndrome measurements and 
recovery operations can reverse errors. However, up to now much less 
is known about the correction of
coherent, unitary errors. Even if a quantum information processor (QIP) is isolated entirely from
its environment and if all 
quantum gates are performed perfectly,
there may still be residual  inter-qubit couplings 
affecting its performance.
Recently, it was demonstrated that static
imperfections, i.e. random inter-qubit couplings which remain 
unchanged during a quantum computation,
restrict the computational capabilities of a many-qubit QIP 
significantly as they cause 
quantum chaos and quantum phase transitions \cite{Shep2}. Furthermore, 
in addition to a usual 
exponential decay such static
imperfections also cause 
a Gaussian decrease of the fidelity with time. 
At sufficiently long times this Gaussian decrease dominates 
the decay of the fidelity thus
limiting significantly the maximum reliable computation 
times of many-qubit QIPs \cite{Benenti01,Shep3}. 

In this Letter a general error correcting method is presented for overcoming these disastrous 
consequences of static imperfections. 
It is based on the repeated random application
of Pauli operators to all the qubits of a QIP.
The resulting random changes of the computational basis together with appropriate compensating changes
of the quantum gates
slow down the rapid Gaussian decay of the fidelity and change it to a
linear-in-time exponential one. 
As a result this Pauli-Random-Error-Correction (PAREC)-method increases significantly 
the maximum time scale of reliable quantum computation.
In addition, neither control measurements nor redundant qubits are required so that all
physical qubits are logical qubits.

In order to put the problem into perspective let us concentrate 
on the quantum algorithm of the quantum tent-map 
as a particular example \cite{Shep3}. One iteration of
this special case of a quantum-rotator-map is governed by the unitary operator
\begin{eqnarray}
\hat{U} &=& e^{-iT\hat{p}^2/(2m \hbar)}e^{-ikV(\hat{x})/\hbar}.
\label{qtent} \end{eqnarray}
It describes the one-dimensional dynamics of a periodically kicked particle of mass $m$.
The operators
$\hat{p}$ and $\hat{x}$ denote momentum and position operators 
and $T$ is the period of the kicks of magnitude $k V(x)$. The dynamics of the particle
is assumed to be confined to the spatial interval $0\leq x\leq l$ with periodic boundary conditions.
The name of this quantum-map originates from 
the force which resembles the form of a tent, i.e.
\begin{eqnarray}
-V^{\prime}(x) = \left\{
\begin{array}{ll}
 (x -\frac{l}{4}),&(0\leq x<l/2)\\
 (\frac{3l}{4} - x),&(l/2\leq x< l).
\end{array}\right. 
\label{tent}
\end{eqnarray}
Due to the periodic boundary  conditions the momentum eigenvalues are given by $p_n = 2\pi\hbar n/l$ with $n\in{\mathbb{Z}}$.
Imposing the 'resonance condition' $T = [m(l/2\pi)^2/\hbar] (2\pi/N)$ with $N \in {\mathbb{N}}$ implies the symmetry 
$\langle p_{n + N} \mid \hat {U}\mid p_{n' + N}\rangle
 = (-1)^{N}\langle p_{n} \mid \hat {U}\mid p_{n'}\rangle$ and $\hat{U}$ decomposes into a direct sum of $N\times N$ matrices \cite{resonance}.
Thus,  for a given value of $N$ the dynamics of the quantum tent-map can be simulated on a quantum computer (QC) with $n_q$ qubits provided
$N=2^{n_q}$. In this case
the unitary operation of Eq.(\ref{qtent}) can be performed with
the help of
$n_g = (9/2) n_q^2 - (11/2)n_q + 4$ universal quantum gates, i.e.
Hadamard-, phase-, controlled-phase-, and controlled-not gates \cite{Shep3}.
The classical limit of the quantum tent-map corresponds to $T\to~0, k\to~\infty$ with $K = kT [l/(2\pi\hbar)]/[m(l/2\pi)^2/\hbar]$ remaining constant. 
In this parameter regime the tent-map exhibits all complex dynamical features characteristic for quantum chaos 
\cite{tent-dynamics}. 

In order to model static imperfections
we assume that the $n_q$ qubits of 
a realistic QC are coupled 
by  random Heisenberg-type
nearest-neighbour interactions as described by the Hamiltonian 
\begin{eqnarray}
\hat{H} &=&\sum_{i=0}^{n_{q}-1} \delta_i \hat{Z}_i 
+ \sum_{i=0}^{n_q-2} J_i(\hat{X}_i\hat{X}_{i+1}+\hat{Y}_i\hat{Y}_{i+1}+
 \hat{Z}_i\hat{Z}_{i+1})\nonumber\\
&&
\label{Heisenberg}
\end{eqnarray}
with
the
Pauli (spin) operators $\hat{X}, \hat{Y}, \hat{Z}$.
The quantities $\delta_i$ and $J_i$ denote the strengths of 
the detuning and of the nearest-neighbour interaction of qubit $i$.
In the case of static imperfections
these quantities
are distributed randomly and homogeneously in the
energy-interval $[-\sqrt{3}\eta, \sqrt{3}\eta]$ and
remain static in time during a quantum computation.

In general, after $t$ iterations of a quantum map
the fidelity, 
defined through the
ideal 
and the perturbed quantum states $\mid \psi(t)\rangle$ and $\mid \psi_{\eta}(t)\rangle$, i.e.
$f=\mid \langle \psi(t) \mid \psi_{\eta}(t)\rangle \mid^2$,
decays according to \cite{Shep3}
\begin{eqnarray}
- {\ln} f(t) &=& \frac{t}{t_c} + \frac{t^2}{t_c t_H}.
\label{decay}
\end{eqnarray}
This relation is valid as long as 
 $\eta$ and $t$ are sufficiently small
so that the fidelity $f$ remains close to unity.
The time scale $t_c$ governing the linear-in-time exponential decay 
is determined by Fermi's Golden rule.
In particular, in recent simulations \cite{Shep3} it was found to be inversely proportional to $(n_q
n_g^2 \eta^2)$.
The second characteristic time scale entering Eq.(\ref{decay}) 
is the Heisenberg time $t_H \approx 2^{n_q} $
which is determined by the dimension of
the $n_q$-qubit Hilbert space.
According to Eq.(\ref{decay}) the linear-in-time exponential 
decay changes to the much more rapid quadratic Gaussian
decay roughly after $t \approx t_H$ iterations.
Such quadratic-in-time Gaussian decays are characteristic 
for coherent dephasing phenomena.
The quadratic fidelity decrease of Eq.(\ref{decay}) 
is drastically limiting the maximum time over which 
a quantum computation can be performed reliably.
Contrary to static imperfections, 
random imperfections which change from gate to gate, e.g., 
lead to a purely
linear-in-time exponential decay of the fidelity \cite{Shep3}.
This suggests the idea that a randomization of static imperfections 
might help to slow down the fidelity decay to a linear-in-time one
thus increasing significantly the reliable computation time.

But how can a randomization of static imperfections be achieved efficiently?
A basic idea of quantum error correction is to
exploit the freedom of choice of the computational basis for an appropriate encoding.
This idea can also be used for an efficient 
randomization of static imperfections by
changing the computational basis 
repeatedly and randomly during a quantum computation. 
However, 
to leave the quantum algorithm unchanged these basis changes have to be compensated
by appropriate transformations of the universal quantum gates.

In order to address this issue
let us concentrate on the particular example of 
the quantum tent-map algorithm and on the static Heisenberg-type
imperfections described by Eq.(\ref{Heisenberg}). 
A convenient way of realizing such random changes of 
the computational basis is to apply repeatedly randomly 
selected Pauli operators to all the
$n_q$ qubits of the QC. 
For this purpose it is advantageous to represent 
the elementary quantum gates of the
quantum algorithm in terms of a special Hamiltonian 
set of universal quantum gates, namely
\begin{eqnarray}
\hat{S}_{\pm X_j}(\Delta \phi) &=& e^{\mp i \hat{X}_j\Delta \phi}, 
\hat{S}_{\pm Z_j}(\Delta \phi) = e^{\mp i \hat{Z}_j\Delta \phi},\nonumber\\ 
\hat{S}_{\pm X_k X_j}(\Delta \phi) &=& e^{\mp i \hat{X}_k \hat{X}_j\Delta \phi}. 
\label{Hamilton}
\end{eqnarray}
The Hamiltonians appearing in the exponents of Eq.(\ref{Hamilton})
are themselves Pauli operators. Therefore,
any unitary transformation $\hat{R}$ originating from Pauli operators either leaves these
Hamiltonians invariant or changes their signs, such as
\begin{eqnarray}
\hat{R}_j \hat{S}_{\pm X_j}(\Delta \phi) \hat{R}_j = 
\left\{
\begin{array}{cc}
\hat{S}_{\pm X_j}(\Delta \phi)&{\rm if}~\hat{R}_j \in \{{\bf 1}_j, \hat{X}_j \}\\
\hat{S}_{\mp X_j}(\Delta \phi)&{\rm if}~\hat{R}_j \in \{\hat{Y}_j, \hat{Z}_j \}. 
\end{array}
\right.
\label{transform}
\end{eqnarray}
Thus, with the help of these Hamiltonian quantum gates 
any change of the computational basis 
originating from 
a randomly selected set 
of Pauli operators 
can be compensated by an appropriate permutation of the universal quantum gates of Eq.(\ref{Hamilton}).

On the basis of these considerations during a quantum computation an efficient correction of static errors 
can be achieved by the PAREC-method in the following 
way (compare with Fig.1).
\begin{figure}
\includegraphics[width=8.6cm]{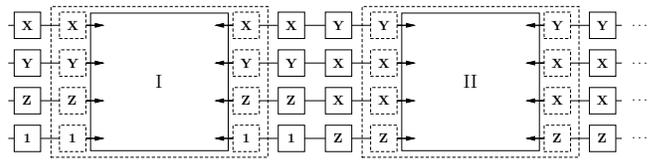}
\caption{The basic idea of the PAREC-method:
The two boxes (full lines) represent two sequences of universal 
quantum gates for $n_q =4$  qubits. Two random sequences 
of Pauli operators $(\hat{X}_1,\hat{Y}_2,\hat{Z}_3,{\bf 1}_4)$ 
and $(\hat{Y}_1,\hat{X}_2,\hat{X}_3,\hat{Z}_4)$ are also indicated.
The unitary Pauli operators outside the dashed boxes (full lines) 
are applied to the qubits whereas the ones inside the dashed boxes (dashed lines)
are taken into account by appropriate permutations
of the elementary quantum gates.
Due to the identities $\hat{X}^2 = \hat{Y}^2 = \hat{Z}^2 = {\bf 1}$ the inserted random sequences
of Pauli operators change the computational basis but leave the ideal quantum algorithm unchanged.}
\end{figure}
In the first step
randomly selected unitary operations 
from the set $\{{\bf 1}, \hat{X},\hat{Y}, \hat{Z}\}$
are applied
to all $n_q$ qubits of the QC, say $(\hat{X}_1, \hat{Y}_2,...,{\bf 1}_{n_q})$.
The information 
about which qubit has been transformed by which Pauli operator 
is stored in a classical memory.
In the second step 
one starts the quantum computation 
by applying a sequence of properly permuted universal quantum gates (compare with
Eqs.(\ref{Hamilton}) and (\ref{transform})) as described by the first dashed box of Fig.1.
This simple permutation 
does not require
any extra significant computational effort.
In the third step 
a second sequence 
of Pauli operators is selected randomly, say
$(\hat{Z}_1,\hat{Y}_2,...,\hat{X}_{n_q})$, and the combined quantum gates
$(\hat{X}_1\hat{Z}_1,\hat{Y}_2 \hat{Y}_2,...,{\bf 1}_{n_q}\hat{X}_{n_q})$ are applied to all the $n_q$ qubits of the QC.
These combined quantum gates are again Pauli operators.
The information about the spin operators of the second selection
is again stored in a classical memory.
Afterwards the second sequence of properly permuted universal quantum gates is performed (second dashed box of Fig.1).
In the subsequent stages of the PAREC-method
these steps are repeated after sequences of universal quantum gates of appropriate lengths $n_{{\rm gef}}$.
The influence of the choice of $n_{{\rm gef}}$ on the error correction will be discussed later (compare with Fig.3).
Finally,
after the application of the last quantum gate the last 
randomly selected sequence of Pauli operators is applied to all qubits.
As apparent from Fig.1 this PAREC-method leaves the ideal quantum algorithm unchanged.
However, the repeatedly applied 
random unitary transformations produced by the Pauli operators 
change the signs
of the parameters $\delta_i$ and $J_i$ of Eq.(\ref{Heisenberg}) 
thus causing a randomization of the static errors. 
As a result we expect a significant improvement 
of the fidelity decay.
\begin{figure}[h]
\includegraphics[width=6.6cm]{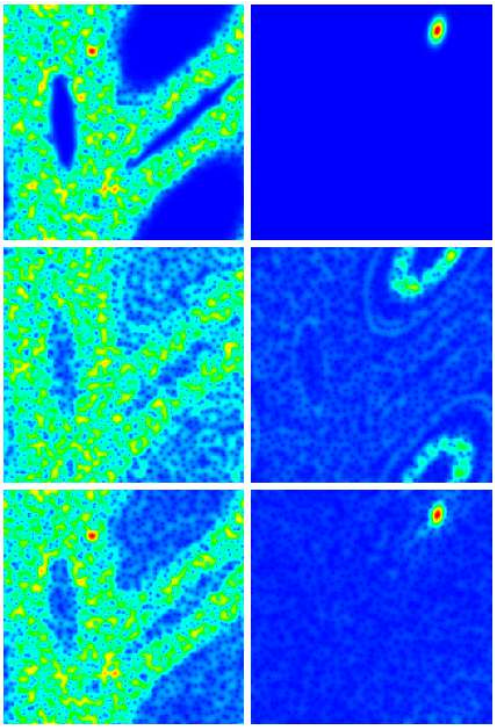}
\caption{(Color) Quantum Poincar\'e sections with Husimi-functions 
at $t = 3000$ in scaled
momentum and position variables $\tilde{y} =  p[(l/(2\pi\hbar)] \in[0,2\pi]$ and $\tilde{x} = x[2\pi/l]\in[0,2\pi]$:
The parameters are $K = 1.7$ and $n_q =10$.
The initially prepared coherent states are centered around $(\pi/4,0)$ (left panel) and
$(5.35,0)$ (right panel).  
First row: ideal dynamics; second row:
static imperfections with $\epsilon = 5\times 10^{-6}$; third row:
PAREC-method applied after each sequence of $n_{{\rm gef}} = 20$ universal quantum gates of Ref. \cite{Shep3}.
The probability density is coded in colors (red/maximum, blue/zero).}
\end{figure}
\begin{figure}[h]
\includegraphics[width=8.6cm]{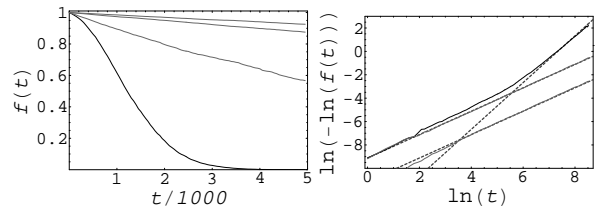}
\caption{Dependence of the fidelity $f(t)$ on the number of iterations: 
Parameters as in the left panel of Fig.2; 
left: static imperfections without error correction, PAREC after each map iteration,
after each $n_{{\rm gef}}= 50$, and after each $n_{{\rm gef}}= 20$ quantum gates (from bottom up); 
right: static imperfections without error correction, with PAREC after each map iteration and after each sequence of $n_{{\rm gef}} = 20$ 
quantum gates (full curves), best fits for linear- and quadratic-in-time decays (dashed curves).}
\end{figure}

The stabilizing properties of the PAREC-method are investigated 
numerically in Figs.2 and 3 where it is applied to
the iterative quantum tent-map.
For this purpose the unitary operator of Eq.(\ref{qtent}) is decomposed into a sequence of 
$\tilde{n}_g$
universal quantum gates of the form
of Eq.(\ref{Hamilton}). This can be achieved in a straightforward way, e.g.,
by corresponding replacements of the  
already known gate decomposition into $n_g$ gates \cite{Shep3}.
The PAREC-randomization is applied after appropriately chosen sequences of
$n_{{\rm gef}}$ quantum gates of this latter decomposition of $n_g$ gates.
The influence of static imperfections is modeled by assuming that these
universal quantum gates are performed instantaneously but that there is a certain time delay between 
any two successive quantum gates during which static imperfections cause errors.
This time delay $\Delta t$ models in an approximate way
the time required for readjustments of the control unit of the QC before it can control the next quantum gate.
As the quantum gates of Eq.(\ref{Hamilton}) involve
different accumulated phases $\Delta \phi$, 
within our model we also assume that quantum gates with larger phases require longer readjustment-times.
Correspondingly, the
readjustment-times $\Delta t_1$ and $\Delta t_2$
after two successive quantum gates with accumulated phases $\Delta \phi_1$ and $\Delta \phi_2$  are related by
$\Delta t_2/\Delta t_1 =\Delta \phi_2/\Delta \phi_1$.
Thus, for a typical sequence of two quantum gates, e.g., the influence of static imperfections
is modeled by
\begin{eqnarray}
\cdots&[e^{-i({\hat{H}\Delta t}/{\hbar})({\Delta \phi_{n+1}}/{\pi})} S_{X_i X_j}(\Delta \phi_{n+1})]&\nonumber\\
&
[e^{-i({\hat{H}\Delta t}/{\hbar})({\Delta \phi_{n}}/{\pi})} S_{X_k}(\Delta \phi_{n})]&\cdots
\end{eqnarray}
with $\Delta t$ denoting the readjustment-time associated with a phase change $\mid \Delta \phi\mid =\pi$.
Correspondingly,
the parameters characterizing the average strength of the influence of the static imperfections between successive quantum gates
are $(\delta_i \Delta t/\hbar)$ and $(J_i \Delta t/\hbar)$
which are selected randomly and uniformly from the interval $[-\sqrt{3}\epsilon, \sqrt{3}\epsilon]$
and which remain constant during the quantum computation.

The ideal dynamics of the quantum tent-map is illustrated in the first row of
Fig.2 where its Husimi-functions \cite{resonance} are plotted after
$t= 3000$ iterations for two initially prepared coherent states  located close to the 
classically unstable (left) and to the classically stable (right) fixed points. 
In the left figure of the first row the initially prepared coherent state spreads almost uniformly over the classically chaotic component
of phase space. Classically inaccessible regions originating from Kolmogorov-Arnold-Moser-tori are also apparent. 
The quantum probability leaking into these regions by quantum tunneling is still negligibly small.
In the right figure of the first row the quantum probability is still concentrated in the region
of the initially prepared coherent state. This reflects the  approximately regular classical dynamics in this part of phase space.
Static imperfections modify these dynamical characteristics significantly, as is apparent from the second row of Fig.2.
Most prominently the influence of quantum tunneling into the classically inaccessible regions of phase space is no longer negligibly small.
Furthermore, the detailed structures in both the regular and the chaotic parts of phase space are modified significantly. 
The corresponding results  of the PAREC-method are depicted in the third row.
Despite the fact that random sequences of Pauli operators are applied only after
each sequence of 
$n_{{\rm gef}} = 20$ gates of Ref. \cite{Shep3}
the corrected
quantum states resemble the ideal case very closely.

The quantitative dependence of the fidelity on the number of iterations
of the quantum tent-map is depicted in Fig.3 for the same parameters as in the left panel of Fig.2.
The initially prepared coherent state of the right panel of Fig.2 yields similar results.
It is apparent that the PAREC-method changes the fidelity decay from quadratic- to linear-in-time, i.e.
$\ln f(t) = - {t}/{t_c}$.
Thus, the time over which quantum computations can be performed reliably is increased significantly.
In addition, one also notices that with increasing numbers of random unitary operations 
the decay rate $1/t_c$ also decreases. In particular, the best fits to the PAREC-results of Fig.3 
together with further numerical studies suggest
a dependence of the form $1/t_c = a \epsilon^2 n_q {n}_g n_{{\rm gef}}$ as long as $1 \ll n_{{\rm gef}}\ll {n}_g$.
Thereby, $n_{{\rm gef}}$ is the effective
number of original gates of Ref. \cite{Shep3}
over which the influence of static imperfections adds up coherently. Our numerical data of Fig.3 give $a\simeq 1$.
Physically speaking this dependence is plausible
on the basis of the following heuristic consideration.
If
the PAREC-method is repeated after each map iteration,
one observes
$n_{{\rm gef}} = {n}_g$ as coherence is destroyed by the random basis changes and by the chaotic dynamics after each iteration.
This result is consistent 
with numerical studies \cite{Shep3}.
In the extreme opposite case in which
the PAREC-method is repeated already after each universal quantum gate one expects $n_{{\rm gef}} = 1$ as coherence is destroyed by
random basis changes already after each gate operation. The above mentioned dependence
interpolates linearly between these two extreme cases. However, due to fluctuations it is expected that these considerations only
apply for sufficiently large values of $n_{{\rm gef}}$.

In summary, a general method for the correction of unitary static inter-qubit errors has been presented.
This PAREC-method
is particularly well suited to
stabilize many-qubit systems against the disastrous effects of static imperfections in arbitrary quantum algorithms.
In contrast to conventional quantum error correcting methods which exploit redundancy
the PAREC-method
does not require any extra qubits so that all physical qubits can be used in an optimal way.

This work is supported in part by the EU IST-FET project EDIQIP
and for DLS by the NSA and ARDA under ARO
contract No. DAAD19-01-1-0553. G. A. also acknowledges support by the DFG (SPP-QUIV).

\end{document}